\documentclass[10pt,journal,compsoc]{IEEEtran}
\ifCLASSOPTIONcompsoc
  \usepackage[nocompress]{cite}
\else
  \usepackage{cite}
\fi

\ifCLASSINFOpdf
  \usepackage[pdftex]{graphicx}
\else
\fi

\usepackage{amsmath}

\ifCLASSOPTIONcompsoc
  \usepackage[caption=false,font=footnotesize,labelfont=sf,textfont=sf]{subfig}
\else
  \usepackage[caption=false,font=footnotesize]{subfig}
\fi

\usepackage{stfloats}

\usepackage{multirow}
\usepackage{hhline}
\usepackage{cleveref}
\crefformat{footnote}{#2\footnotemark[#1]#3}

\usepackage{amsthm}

\newtheorem{definition}{Definition}

\usepackage{pgfplots,pgfplotstable}
\pgfplotsset{compat=newest}
\usetikzlibrary{patterns,arrows,automata,positioning,calc}
\usepgfplotslibrary{colorbrewer,statistics}

\usepackage{acronym}

\acrodef{DVFS}{dynamic voltage and frequency scaling}
\acrodef{QoS}{quality-of-service}
\acrodef{NoC}{network-on-chip}
\acrodef{SEU}{single event upset}
\acrodef{DMR}{dual modular redundancy}
\acrodef{TMR}{triple modular redundancy}
\acrodef{CRC}{cyclic redundancy check}
\acrodef{EDC}{error-detecting code}
\acrodef{ECC}{error-correcting code}
\acrodef{FMEA}{failure mode and effect analysis}
\acrodef{MPSoC}{multi\-processor system-on-chip}
\acrodefplural{MPSoC}{multi\-processor systems-on-chip}%
\acrodef{ARQ}{automatic repeat request}
\acrodef{BER}{bit error rate}
\acrodef{DMA}{direct memory access}
\acrodef{MTTF}{mean time to failure}
\acrodef{FIT}{failure in time}
\acrodef{SIL}{Safety Integrity Level}
\acrodef{FPGA}{field-programmable gate array}
\acrodef{ASIC}{application-specific integrated circuit}
\acrodef{OS}{operating system}
\acrodef{RTOS}{real-time operating system}
\acrodef{FMEA}{failure mode and effects analysis}
\acrodefplural{FMEA}{failure mode and effects analyses}%
\acrodef{API}{application programming interface}
\acrodef{WCET}{worst-case execution time}
\acrodef{CPA}{Compositional Performance Analysis}
\acrodef{flit}{flow control unit}
\acrodef{RTE}{runtime environment}
\acrodef{ASIL}{Automotive Safety Integrity Level}

\acrodef{IPF}{Information Processing Factory}
\acrodef{SC}{system controller}
\acrodef{BEC}{best-effort controller}
\acrodef{TAL}{Trace Abstraction Layer}
\acrodef{LCT}{Learning Classifier Table}
\acrodef{OR}{Operating Region}
\acrodef{NOR}{Next Operating Region}
\acrodef{COR}{Current Operating Region}
\acrodef{OP}{Operating Point}

\usepackage{color}
\usepackage{tabularx}
\usepackage{enumitem}

\begin{document}
%
\title{The Information Processing Factory:\\ Organization, Terminology, and Definitions}
%
%
%
%

\author{Eberle~A.~Rambo,
        Bryan~Donyanavard,
        Minjun~Seo,
        Florian~Maurer,
        Thawra~Kadeed,
        Caio~B.~de~Melo,
        Biswadip~Maity,
        Anmol~Surhonne,
        Andreas~Herkersdorf,
        Fadi~Kurdahi,
        Nikil~Dutt,
        and~Rolf~Ernst
\IEEEcompsocitemizethanks{\IEEEcompsocthanksitem E. A. Rambo, T. Kadeed, and R. Ernst are with Institute of Computer and Network Engineering, Technische Universit\"{a}t Braunschweig, Braunschweig, Germany.\protect\\
E-mails: rambo@ida.ing.tu-bs.de, kadeed@ida.ing.tu-bs.de, and ernst@ida.ing.tu-bs.de
\IEEEcompsocthanksitem B. Donyanavard, M. Seo, C. B. de Melo,
        B. Maity, N. Dutt, and F. Kurdahi are with Center for Embedded and Cyber-physical Systems (CECS), University of California, Irvine, USA.\protect\\
E-mails: bdonyana@uci.edu, minjun.seo@uci.edu, cbatista@uci.edu, maityb@uci.edu, dutt@ics.uci.edu, and kurdahi@uci.edu
\IEEEcompsocthanksitem F. Maurer, A. Surhonne, and A. Herkersdorf are with Chair for Integrated Systems, Technische Universit\"{a}t M\"{u}nchen, M\"{u}nchen, Germany.\protect\\
E-mails: flo.maurer@tum.de, anmol.surhonne@tum.de, and herkersdorf@tum.de}
}

\markboth{}%
{}

\maketitle


%
\IEEEpeerreviewmaketitle

\acresetall


\IEEEraisesectionheading{\section{Introduction}\label{sec:introduction}}

\IEEEPARstart{T}{he} \ac{IPF} project has recently introduced the abstraction of complex architectures as self-aware information processing factories \cite{dutt2016conquering, sadighi2018design}.
These factories consist of a set of highly configurable resources, e.g., processing elements and interconnects, whose use is monitored, planned, and configured during runtime.
Managing a factory involves multiple facets, such as efficiency, availability, reliability, integrity, and timing.
\ac{IPF} conquers the complexity of managing facets in digital systems by hierarchically decomposing the challenges and addressing them with different co-existing entities in the factory.

This paper introduces the organization, terminology, and definitions of \ac{IPF}.
Section~\ref{sec:factory} describes \ac{IPF}'s five-layer hierarchical organization and a system configuration framework that enable self-awareness, self-diagnosis, self-organization, and self-optimization in mixed-critical real-time embedded systems.
Section~\ref{sec:imminenthazards} defines imminent hazards, which are threats to the system operation that can be handled by \ac{IPF} before they can cause system failure.
Section~\ref{sec:conclusion} concludes the paper.


\section{IPF's 5-layer Organization}\label{sec:factory}
\ac{IPF} is a metaphor for self-aware, self-organizing, mixed-critical systems.
\ac{IPF} provides an infrastructure for system introspection and reflective behavior, which is the foundation for computational self-awareness.
Computational self-awareness is the ability of a computing system to recognize its own state, possible actions and the result of these actions on itself, its operational goals, and its environment, thereby empowering the system to become autonomous \cite{jantsch2017self}.
An \ac{IPF} system consists of a set of highly configurable resources whose use is planned, configured, monitored, and optimized at runtime.
The system resources are allocated to the execution of mixed-critical workloads.
Thus, not only must the factory manage resources and workload at runtime, it must do so while ensuring that the requirements of the safety-critical functions of the workload are not violated.
\ac{IPF} addresses that challenge with two levels of awareness: a first level with local, autonomous actions and a second level with a global view of the system.
Together, they ensure the adaptability, safety, and dependability of the mixed-critical system throughout its execution.

\begin{figure*}[ht]
\centering
\includegraphics[width=1.4\columnwidth]{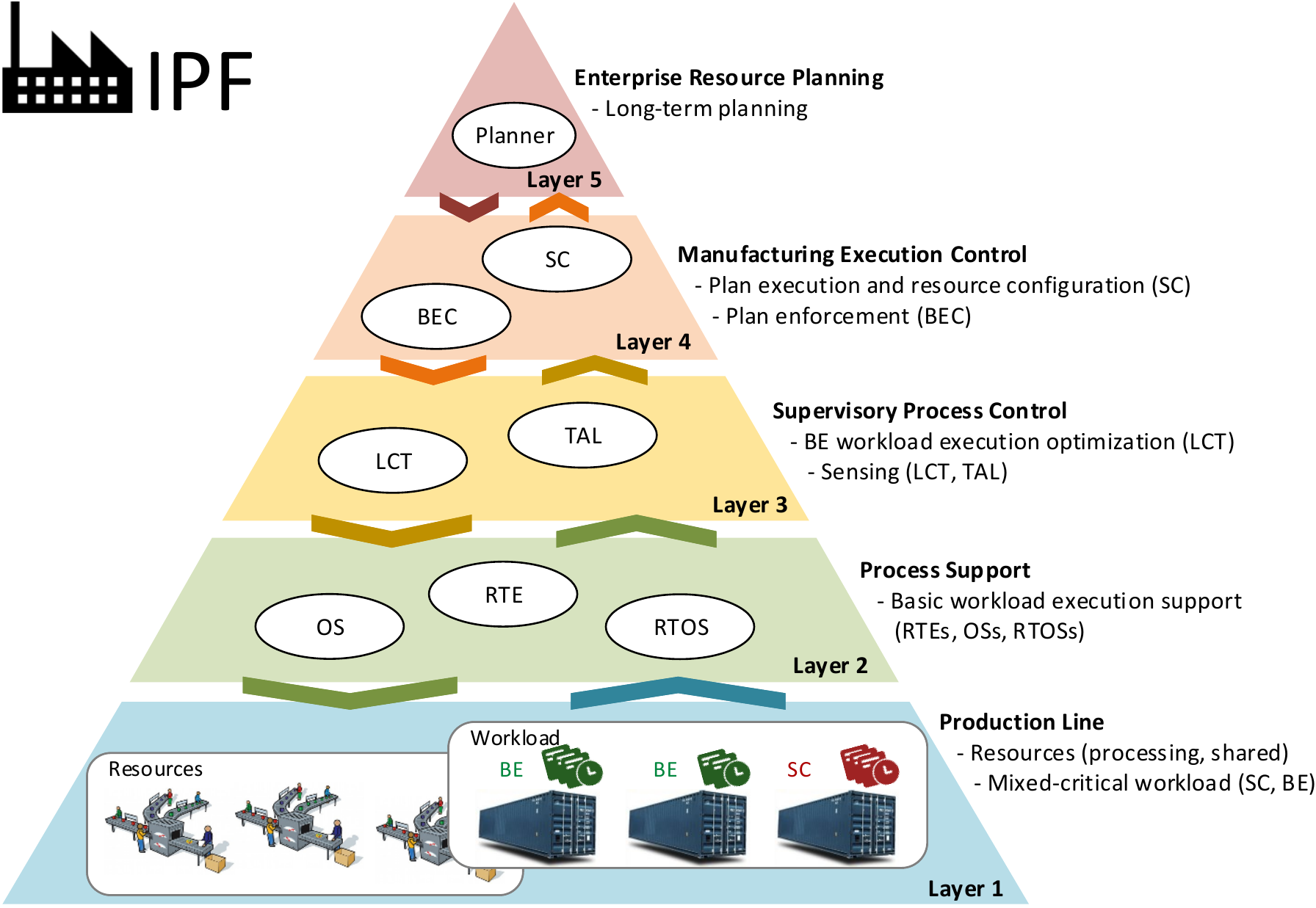}
\caption{\ac{IPF}'s five-layer organization.}
\label{fig:factory:overview}
\end{figure*}

Resembling a factory\footnote{Terminology inspired by enterprise control systems \cite{isa95part1}.}, \ac{IPF} is organized in five layers, as illustrated in Figure~\ref{fig:factory:overview}.
The workload execution occurs in the \emph{production line} (layer 1), which contains the system resources and the mixed-critical workload.
The workload executes within the infrastructure and execution model of the \emph{process support} (layer 2), which provides basic execution support such as the \acp{OS}, \acp{RTOS}, and the \acp{RTE}.
The resources' statuses are monitored and the workload execution is optimized by the \emph{supervisory process control} (layer 3), which acts locally and autonomously within boundaries specified by the layers above.
The \emph{manufacturing execution control} (layer 4) is responsible for enforcing safe system configurations by globally monitoring, assessing risks and controlling the layers below, under the guidance of the top layer.
The \emph{enterprise resource planning} (layer 5) is responsible for long-term planning of \ac{IPF}.
It plans future proactive and reactive actions, taking into account the operating conditions of the system, assessing risks and impacts of short-term factors such as error rates, energy consumption, workload variations; and long-term factors such as aging, energy constraints, and changes in the workload, \ac{QoS} goals, and non-functional constraints.
Note that, in the absence of the top three layers, layers 1 and 2 compose the original non-self-aware system.

\subsection{Production Line (Layer 1)}\label{sec:factory:assembly}
The lowest layer of \ac{IPF} is responsible for the workload execution.
Depicted in Figure~\ref{fig:factory:overview}, the production line consists of the system resources and the mixed-critical workload.

The mixed-critical workload consists of a best-effort (BE) component and a safety-critical (SC) component.
The BE workload is characterized by its goals, and the SC workload is characterized by its non-functional requirements.
The BE workload has application-specific \ac{QoS}, such as achieved throughput, and also has system constraints such as power budget.
The SC workload has requirements, such as period, \ac{WCET}, deadline, maximum downtime, maximum \ac{FIT}, and data consistency.
Different levels of criticality \cite{ernst2016mcs} as defined in safety standards \cite{iso26262, iec61508, do254} are also supported.
The five usual levels of criticality -- e.g., \acp{ASIL} A through D plus QM, are captured by the different requirements of the workload.
For simplicity and without loss of generality, throughout the paper we refer to the two representative levels: BE and SC.

Processing resources are highly configurable and characterized by their properties -- e.g., operating frequencies, power consumption, temperature, and error rates.
They can consist of a single or multiple cores in a cluster.
Processing resources can be used for executing either BE or SC workload.
Therefore, \ac{IPF} requires them to have a \emph{safety-critical mode} where the execution is deterministic, predictable, and enables minimum performance guarantees for applications that require it -- i.e., the SC workload.
For BE workload execution, processing resources can enable non-predictable features, such as caches in the memory hierarchy and \ac{DVFS}.
Shared resources must provide predictable and deterministic service and ensure sufficient independence\footnote{\label{note_iso}Also known as freedom from interference \cite{iso26262}.} between different criticalities \cite{iec61508}, thereby enabling minimum performance guarantees for the SC workload.
The resources are configured by the manufacturing execution control (layer 3), which enables and disables configurable features according to the executing workload.

\begin{figure}[h]
\centering
\includegraphics[width=0.75\columnwidth]{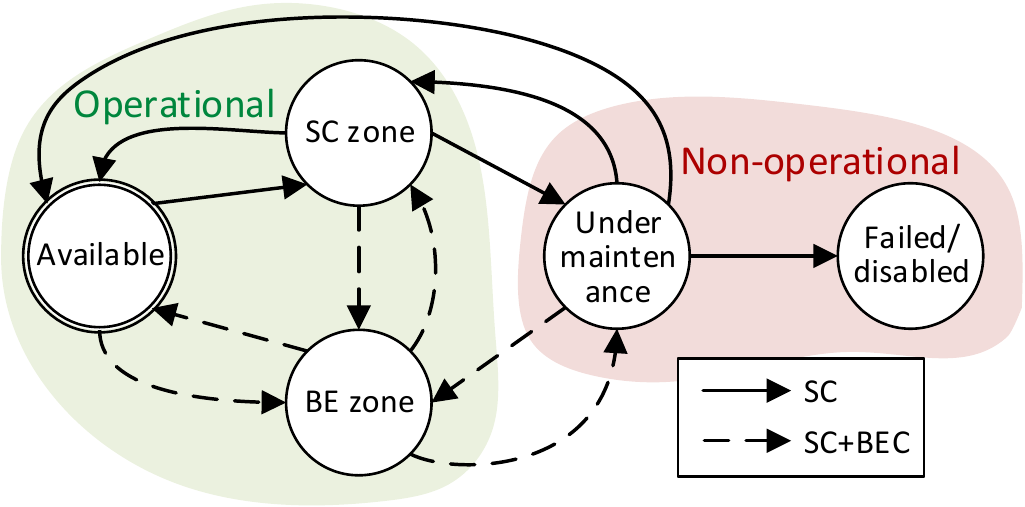}
\caption{Resource states: resource planning is done in layer 5, resource management (state transitions) is done in layer 4.}
\label{fig:factory:resourcestates}
\end{figure}

Processing resources in \ac{IPF} have five different states and two superstates, as illustrated in Figure~\ref{fig:factory:resourcestates}.
A processing resource can either be operational or non-operational.
When operational, a processing resource can be allocated to the SC workload execution or the BE workload execution.
In the former case, the respective processing resource belongs to the \emph{SC zone}, and in the latter, to the \emph{BE zone}.
When non-operational, e.g., due to errors, the processing resource can be either under maintenance or failed.
If a processing resource is under maintenance, \ac{IPF} attempts to find a configuration under which the resource can still be employed.
When failed, the resource is no longer usable.
The resource planning is done in layer 5.
The resource management is done at layer 4, which is responsible for the different transitions.
Shared resources are managed similarly and serve both SC and BE zones.

\subsection{Process Support (Layer 2)}\label{sec:factory:level1}
The second layer of \ac{IPF} is responsible for basic workload execution support and the execution model.
The process support comprises elements such as \acp{OS}, \acp{RTOS}, and \acp{RTE}, as illustrated in Figure~\ref{fig:factory:overview}.
Notice that the lowest two layers alone comprise a regular mixed-critical system without any self-* properties.

For modularity and to enable predictable dynamicity at runtime, the execution model of \ac{IPF} is based on containers.
A container encapsulates a (sub)set of either SC or BE workload, and is referred to as an \emph{SC container} or a \emph{BE container}, respectively.
A container also includes an \ac{RTE} with \ac{OS} or \ac{RTOS}, and a software stack.
Each container is mapped to a processing resource, which is associated with a single container -- i.e., there exists a strict one-to-one mapping between containers and processing resources.
The mapping of workload to containers and the mapping of containers to resources are specified in \acp{OR}, to be introduced in Section~\ref{sec:factory:opsors}.
At runtime, the workload can be redistributed among different containers, and containers can be moved between resources, e.g., with workload balancing and migration techniques.
That can be carried out by entities in layers 4 and 5 by means of transitions between \acp{OR}.

\subsection{Operating Regions and Operating Points}\label{sec:factory:opsors}
Before advancing to the next layers, we define the concepts of \acp{OR} and \acp{OP}.
The vision for \acp{OP} was introduced in \cite{sadighi2018design}.
Here, we apply the concept in a framework for configuring and managing the system.
The framework consists of \acp{OR} and \acp{OP}.
An \ac{OR} is a configuration of the system with room for small changes.
Small changes are carried out by changing the \acp{OP} inside it.
Larger changes are carried out by changing the \ac{OR}.

An \ac{OR} represents a configuration of the system where the mixed-critical workload (including goals and requirements), its mapping to BE and SC containers, the mapping of containers to resources, and the configuration of the shared resources are fixed.
In an \ac{OR}, the configuration of the containers and the associated processing resources (in the BE zone) can be varied.
What can be varied depends on the specific instance of \ac{IPF} and on the underlying hardware.
How much it can be varied, i.e., the configuration range, is determined at runtime in layer 5, introduced in Section~\ref{sec:factory:director}.
Both what can be varied and how much it can be varied are specified in the \ac{OR}.
The intuition behind \acp{OR} is that they represent valid system configurations where the system is predictable and safe for executing the SC workload, while still providing safely bounded flexibility for local optimizations of the execution of BE workload.

\begin{figure*}[t]
\centering
\captionsetup[subfloat]{captionskip=2pt}
\subfloat[Defining \acp{OR} and \acp{OP}]{%
\includegraphics[width=1.28\columnwidth]{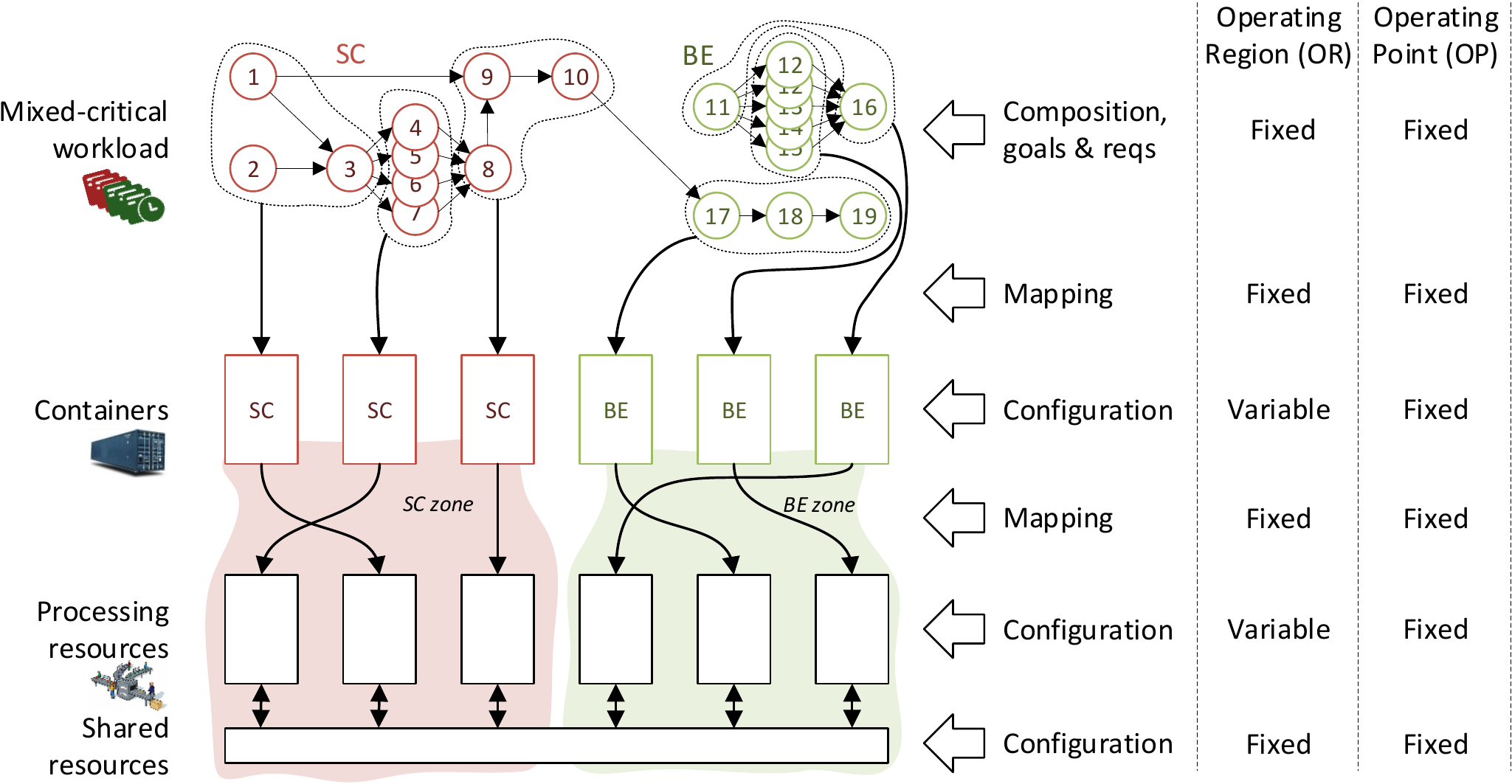}
\label{fig:factory:operatingpoints:def}}\hspace{8pt}
\subfloat[\acp{OR}, \acp{OP}, and events]{%
\includegraphics[width=0.57\columnwidth]{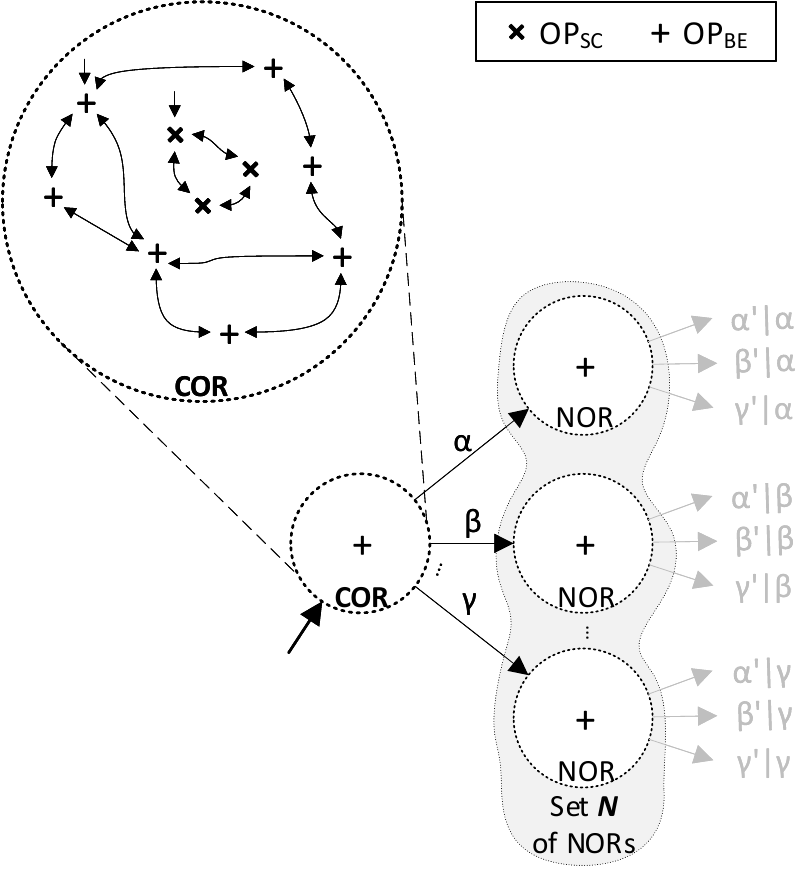}
\label{fig:factory:operatingpoints:regions}}
\caption{Operating regions: \acf{COR} and the set of \acf{NOR}, triggered by different events ($\alpha$, $\beta$, and $\gamma$). An operating region can have multiple valid \acfp{OP}.}
\label{fig:factory:operatingpoints}
\end{figure*}

An \ac{OP} is a specific configuration within an \ac{OR}.
It can also be decomposed in two components, and be defined as a pair: OP $=($OP$_{SC}, $OP$_{BE})$.
The OP$_{BE}$ represents a specific configuration of the BE zone, and OP$_{SC}$ represents a specific configuration of the SC zone.
The concept is illustrated in the top left quadrant of Figure~\ref{fig:factory:operatingpoints:regions}, where there are multiple operating points within a region.
The intuition behind \acp{OP} is that they represent a specific, valid configuration at any given point in time within an \ac{OR}.

As illustrated in Figure~\ref{fig:factory:operatingpoints:regions}, the system starts at an initial, valid \ac{OR}, named \ac{COR}.
A number of autonomous actions can be performed locally by the system, which move the \ac{OP} around inside the \ac{COR}.
Whenever an event occurs in \ac{IPF} that requires bigger changes to the configuration, \ac{IPF} handles it by transitioning to a new \ac{OR}, named \ac{NOR}.
A suitable \ac{NOR} is chosen from a set $N$ of \acp{NOR} and \ac{IPF} then reconfigures the system according to the selected \ac{NOR}, which becomes the \ac{COR}.
Then, the set $N$ of \acp{NOR} becomes empty and new, valid \acp{NOR} must be created and added to~$N$.

Events can be triggered due to different reasons, such as:
\begin{itemize}
    \item anticipated violation of the requirements of the safety-critical workload, i.e., a hazard;
    \item pursuit of long-term optimization goals;
    \item changes in the workload, its goals and requirements;
    \item changes in the environmental conditions;
    \item changes in the operating conditions.
\end{itemize}
Events are generated by the entities in layers 3 and 4 and will be introduced in the respective layers.

\subsection{Supervisory Process Control (Layer 3)}\label{sec:factory:level2}
The supervisory process control is responsible for monitoring and autonomously optimizing the workload execution.
It is also responsible for gathering useful information about the production line that supports the upper layers' long-term planning and execution.
Supervisory process control components carry out actions that directly modify the configuration of the system by changing its \ac{OP}.
As illustrated in Figure~\ref{fig:factory:overview}, the layer comprises the \ac{IPF} infrastructure components \ac{TAL} and \ac{LCT}.

\ac{TAL} is responsible for monitoring the system for errors.
\ac{TAL} is a source of events that trigger \ac{OR} transitions.
\ac{TAL} performs runtime verification based on processor tracing.
It checks the execution of the workload against contracts (system requirements) described as Timed Automata (TA) models.
The contracts are loaded into \ac{TAL}, which continuously monitors the system at runtime.
\acp{TAL} operate in both BE and SC containers.

The \ac{LCT} is responsible for optimizing the execution of the BE workload towards achieving goals.
\acp{LCT} \cite{zeppenfeld2008learning} are rule-based reinforcement learning engines that explore and optimize configurations within the \ac{COR}.
They operate only within BE containers.
\acp{LCT} collect periodic sensor data to update the fitness of rules and determine the action for the next period.
First, based on the effect of the previous action toward achieving an objective, the \ac{LCT} updates the rule fitness for the previous period using a version of Q-learning. 
Second, based on the current state and rule fitnesses, the \ac{LCT} applies an action to configure the system for the upcoming period by changing the \ac{OP} within the \ac{COR} (cf. Section~\ref{sec:factory:opsors}).
\acp{LCT} and \acp{TAL} are configured and maintained by layer 3.

\subsection{Manufacturing Execution Control (Layer 4)}\label{sec:factory:manager}
In the factory analogy, the manufacturing execution control is responsible for the global monitoring, risk assessment, and control of the system.
It monitors the layers below with the support of layer 3 and controls them by means of the \acp{OR} provided by the enterprise resource planning (layer 5).
Changes to the system configuration initiated by this layer are seen as big changes and are realized with transitions from a \ac{COR} to a \ac{NOR} (cf. Section~\ref{sec:factory:opsors}).

As illustrated in Figure~\ref{fig:factory:overview}, the layer comprises two entities: the \emph{best-effort controller} (BEC) and  the \emph{system controller} (SC). \acused{BEC}\acused{SC}
\ac{SC} monitors and controls the safety-critical part of the system (the SC zone) as well as the shared resources according to the \ac{COR}.
\ac{BEC} monitors and controls the best-effort part of the system, the BE zone, according to the \ac{COR}.
\ac{SC} coordinates the control of the BE zone with \ac{BEC}, but for safety, \ac{SC} has ultimate control over the entire system.

The \ac{SC} is responsible for configuring the system according to the \ac{COR}.
It configures the resources and the entities in the lower layers of \ac{IPF}, loads the SC and BE containers onto the respective resources, and configures the shared resources as specified by the \ac{COR}.
The \ac{BEC} is responsible for configuring the resources and entities in the BE zone according to the \ac{COR}, and ensures that the autonomous actions carried out in the lower layers are within the specification of the \ac{COR}.

Both entities globally monitor the system, assess risks, and proactively and reactively act to changes in the system or environment.
These proactive and reactive measures in the system are triggered by events (cf. Section~\ref{sec:factory:opsors}), which trigger big changes in the configuration of the system.
Events are generated either in layer 3 or in this layer.
\ac{SC} monitors the system for changes in the environmental or operating conditions that impact the execution of the SC workload.
Similarly, \ac{BEC} monitors the BE zone for changes in the system or environment that impacts the execution of the BE workload.
Events refer to, for example, changes in the BE workload and its goals, changes in the SC workload and its requirements, or the failure of a processing resource, which will be reactively handled in \ac{IPF}.
An event can also refer to a predicted change that will be proactively handled by \ac{IPF}, such as the imminent failure of a processing resource.
Because changes in the global system configuration impact the SC workload execution, \ac{SC} is responsible for handling all events, including events that occur in the BE zone, which are forwarded by \ac{BEC} to \ac{SC}.
Handling the event means transitioning from the \ac{COR} to a suitable \ac{NOR} associated with that event.

Transitions from the \ac{COR} to \acp{NOR} are carried out by \ac{SC} with the collaboration of \ac{BEC}.
Such transitions can include changes to the configuration of a single resource or can include a complete change in the mapping of workload to containers and their mapping to resources.
During a transition, in addition to the above-described configuration responsibilities, resources can be added and removed from the SC and BE zones.
The management of a resource's state by the \ac{SC} and \ac{BEC} controllers is illustrated in Figure~\ref{fig:factory:resourcestates}, where solid-arrow changes to the resource state involve only \ac{SC}, and dashed-arrow transitions involve both \ac{SC} and \ac{BEC}.
When the SC and BE zones are resized, i.e., resources are added to or removed from a zone, a controller appropriately releases its resources before handing them over to the other one -- e.g., \ac{BEC} removes a resource from the BE zone before handing it over to \ac{SC} to be added to the SC zone.
Notice that the transition of a resource from BE zone to SC zone in Figure~\ref{fig:factory:resourcestates} is timing critical.
Therefore, \ac{SC} is allowed to forcefully execute that transition without \ac{BEC}, in case the latter takes too long to release it, in order to make the execution of the SC workload independent of the execution of the BE workload (sufficient independence).
Shared resources are configured by \ac{SC} since they must comply with the highest levels of criticality \cite{iso26262}, and they must be reconfigured before any BE containers can resume execution in order to ensure sufficient independence and prevent unexpected interference on the execution of the SC workload.
Note that the time to transition between \acp{OR} varies depending on the amount of reconfiguration involved.

When an event occurs for which there is no associated \ac{NOR}, the reaction of \ac{IPF} depends on the event.
The possible scenarios are summarized in Table~\ref{tab:IPF_failure_scenarios}.
In case the event is related to the execution of the SC workload, the system must signal its failure before the non-functional properties of the workload are violated.
The failure is either reported immediately if the event concerns a reactive measure, or it can be deferred to when the failure actually occurs if the event concerns a proactive measure.
If the event is related to the execution of BE workload, the system continues operating albeit with limited \ac{QoS}, which can be optionally reported.
Such scenarios can be triggered either because there are actually no more valid \acp{NOR} for that event, e.g., due to a number of resource failures leading to an insufficient number of resources, or because a valid \ac{NOR} has not been created yet (but eventually will), e.g., due to a quick succession of events.

\begin{table}[ht]
\centering
\caption{Failure scenarios in \ac{IPF} where an event concerning either BE or SC workload has no associated \ac{NOR} with a reactive or proactive measure.}
\label{tab:IPF_failure_scenarios}
\begin{tabular}{l|c|c|}
\cline{2-3}
\textbf{}                                  & \textbf{Reactive}      & \textbf{Proactive}           \\ \hline
\multicolumn{1}{|l|}{\textbf{BE workload}} & \multicolumn{2}{c|}{Limited QoS (optional)} \\ \hline
\multicolumn{1}{|l|}{\textbf{SC workload}} & Failure report         & Deferred failure report      \\ \hline
\end{tabular}
\end{table}

\subsection{Enterprise Resource Planning (Layer 5)}\label{sec:factory:director}
Finally, the enterprise resource planning is responsible for the long-term planning of the system.
That is, developing the future configurations of the system in the form of \acp{NOR}.
The planning is supported by system information supplied by layers 3 and 4, including the resources and their current operating conditions.
Planning also considers the system's current and previous \acp{OR}; current and previous operating conditions; the workload, its \ac{QoS} goals and non-functional constraints; short-term and long-term factors, such as error rates, energy consumption, aging, and energy constraints; and the different events that may occur at runtime.

The \emph{planner} is the main entity of this layer.
Its main responsibility is to create and maintain the set $N$ of \acp{NOR}.
$N$ is modified when \ac{IPF} transitions to a \ac{NOR}, which requires a new set of \acp{NOR}, and $N$ is therefore emptied; and when the planner creates a new, valid \ac{NOR}, in which case a new \ac{NOR} is added to $N$.
The planner also defines the valid \acp{OP} in \acp{OR} -- i.e., the configuration ranges in an \ac{OR} within which \ac{IPF}'s local autonomous actions and optimization in layer 3 can operate.
That is required due to possible coupling between system resources.
For example, physical temperature coupling, where the high temperature of a processing resource can affect neighboring resources and increase their error rates.

An \ac{OR} is a valid configuration range of the system, and therefore the planner only includes a new \ac{NOR} in the set $N$ if it meets all non-functional requirements of the SC workload.
That includes checking the \acp{OR} and their \acp{OP} with, e.g., system-level performance analysis tools such as \ac{CPA} \cite{henia2005system}.
The planner can take the \ac{QoS} goals of the BE workload into consideration, but there is no guarantee that the goals will be met in a given \ac{OR}.

The transitions between different \acp{OR} are triggered by different events.
Independent of the event, the planner must consider the cost of transitioning between different \acp{OR}.
Transitions can involve the remapping of workload to containers and the remapping of containers to resources.
Remapping requires moving code and data and therefore impacts the response time of the executing workload, which can lead to system-level timing violations (deadline misses).
Thus, the planner must also check for system-level timing and safety violations of the transition from the \ac{COR} to the \ac{NOR} before the \ac{NOR} can be included in the set $N$.


\section{Imminent Hazards}\label{sec:imminenthazards}

\ac{IPF} goes beyond conventional hardware faults by addressing \emph{imminent hazards}.
Intuitively, a hazard is a consequence of the system failing to function as expected.
From ISO~26262, a failure is the termination of the ability of the system or sub-system to perform a function as required \cite{iso26262}.
A hazard is a potential source of harm caused by the malfunctioning behaviour of the system, i.e., a system failure \cite{iso26262}.
An imminent hazard can then be defined as follows:

\begin{definition}\label{definition:imminenthazard}
\emph{Imminent hazard:} an increased risk of future errors that can lead to system failure and, therewith, a hazard.
\end{definition}

An imminent hazard can be caused by physical causes, environmental conditions, or operating conditions.
They can be caused, e.g., by the imminent failure of a resource due to an impending permanent fault.
When a permanent fault is close to occur due to aging processes, increasing error rates and intermittent faults are observed \cite{constantinescu2003trends}.
In this case, imminent hazards must be distinguished from latent faults and regular fault occurrences.
Imminent hazards can also be caused by error rates for which the system is not dimensioned to tolerate.


\section{Conclusion}\label{sec:conclusion}

This paper described \ac{IPF}'s five-layer hierarchical organization and system configuration framework, and defined the concept of imminent hazards.


%

%

\ifCLASSOPTIONcompsoc
  \section*{Acknowledgments}
\else
  \section*{Acknowledgment}
\fi

We acknowledge financial support from the following: NSF Grant CCF-1704859; DFG Grants ER168/32-1 and HE4584/7-1.

\ifCLASSOPTIONcaptionsoff
  \newpage
\fi



\bibliographystyle{IEEEtran}
\bibliography{references}

\begin{thebibliography}{10}
\providecommand{\url}[1]{#1}
\csname url@samestyle\endcsname
\providecommand{\newblock}{\relax}
\providecommand{\bibinfo}[2]{#2}
\providecommand{\BIBentrySTDinterwordspacing}{\spaceskip=0pt\relax}
\providecommand{\BIBentryALTinterwordstretchfactor}{4}
\providecommand{\BIBentryALTinterwordspacing}{\spaceskip=\fontdimen2\font plus
\BIBentryALTinterwordstretchfactor\fontdimen3\font minus
  \fontdimen4\font\relax}
\providecommand{\BIBforeignlanguage}[2]{{%
\expandafter\ifx\csname l@#1\endcsname\relax
\typeout{** WARNING: IEEEtran.bst: No hyphenation pattern has been}%
\typeout{** loaded for the language `#1'. Using the pattern for}%
\typeout{** the default language instead.}%
\else
\language=\csname l@#1\endcsname
\fi
#2}}
\providecommand{\BIBdecl}{\relax}
\BIBdecl

\bibitem{dutt2016conquering}
N.~Dutt, F.~J. Kurdahi, R.~Ernst, and A.~Herkersdorf, ``Conquering {MPSoC}
  complexity with principles of a self-aware information processing factory,''
  in \emph{Proceedings of the 11th IEEE/ACM/IFIP International Conference on
  Hardware/Software Codesign and System Synthesis (CODES+ISSS)}, ser. CODES'16,
  Pittsburgh, Pennsylvania, Oct 2016.

\bibitem{sadighi2018design}
A.~Sadighi, B.~Donyanavard, T.~Kadeed, K.~Moazzemi, T.~M{\"u}ck, A.~Nassar,
  A.~M. Rahmani, T.~Wild, N.~Dutt, R.~Ernst, A.~Herkersdorf, and F.~J. Kurdahi,
  ``Design methodologies for enabling self-awareness in autonomous systems,''
  in \emph{Proceedings of Design, Automation and Test in Europe Conference
  (DATE’18)}, March 2018.

\bibitem{jantsch2017self}
A.~{Jantsch}, N.~{Dutt}, and A.~M. {Rahmani}, ``Self-awareness in systems on
  chip-- a survey,'' \emph{IEEE Design Test}, vol.~34, no.~6, pp. 8--26, Dec
  2017.

\bibitem{isa95part1}
``{ANSI/ISA}-95.00.01-2010 ({IEC} 62264-1 mod) - enterprise-control system
  integration - part 1: Models and terminology,'' International Society of
  Automation, 2010.

\bibitem{ernst2016mcs}
R.~{Ernst} and M.~{Di Natale}, ``Mixed criticality systems�a history of
  misconceptions?'' \emph{IEEE Design Test}, vol.~33, no.~5, pp. 65--74, Oct
  2016.

\bibitem{iso26262}
``{ISO}~26262: Road vehicles -- functional safety,'' International Standards
  Organization, 2018.

\bibitem{iec61508}
``{IEC} 61508: Functional safety of electrical/electronic/programmable
  electronic safety-related systems, ed.2.0,'' International Electrotechnical
  Commission, 2010.

\bibitem{do254}
``{DO}-254: Design assurance guidance for airborne electronic hardware,'' RTCA
  Incorporated, 2000.

\bibitem{zeppenfeld2008learning}
J.~Zeppenfeld, A.~Bouajila, W.~Stechele, and A.~Herkersdorf, ``Learning
  classifier tables for autonomic systems on chip.'' \emph{GI Jahrestagung
  (2)}, vol. 134, pp. 771--778, 2008.

\bibitem{henia2005system}
R.~Henia, A.~Hamann, M.~Jersak, R.~Racu, K.~Richter, and R.~Ernst, ``System
  level performance analysis--the symta/s approach,'' \emph{IEE
  Proceedings-Computers and Digital Techniques}, vol. 152, no.~2, pp. 148--166,
  2005.

\bibitem{constantinescu2003trends}
C.~Constantinescu, ``Trends and challenges in vlsi circuit reliability,''
  \emph{IEEE micro}, vol.~23, no.~4, pp. 14--19, 2003.

\end{thebibliography}
\end{document}